%% file: 0.main.tex
\documentclass[manuscript,anonymous=false]{acmart} 

\AtBeginDocument{%
  \providecommand\BibTeX{{%
    \normalfont B\kern-0.5em{\scshape i\kern-0.25em b}\kern-0.8em\TeX}}}

\setcopyright{acmcopyright}
\copyrightyear{2018}
\acmYear{2018}
\acmDOI{10.1145/1122445.1122456}

\usepackage[utf8]{inputenc}
\usepackage[T1]{fontenc}
\DeclareUnicodeCharacter{1EC7}{\d{\^e}}

\usepackage{booktabs} 
\usepackage{url}

\usepackage{caption}
\usepackage[all]{nowidow}
\usepackage{wrapfig}
\usepackage{array}
\usepackage{arydshln}
\usepackage{tabularx}
\usepackage{multirow}
\usepackage{arydshln}
\newcolumntype{L}[1]{>{\raggedright\let\newline\\\arraybackslash\hspace{0pt}}m{#1}}
\newcolumntype{C}[1]{>{\centering\let\newline\\\arraybackslash\hspace{0pt}}m{#1}}
\newcolumntype{R}[1]{>{\raggedleft\let\newline\\\arraybackslash\hspace{0pt}}m{#1}}

\usepackage{wrapfig,lipsum,booktabs} 

\def\authnotes{1}
\newcounter{notectr}[section]
\newcommand{\thenote}{\thesubsection.\arabic{notectr}\refstepcounter{notectr}}

\newcommand{\note}[2]{$\ll$#1~\thenote: #2$\gg$}
\newcommand{\cnote}[1]{\ifnum\authnotes=1 \textcolor{blue}{\note{Comment:}{#1}}\fi}

\copyrightyear{2026}
\acmYear{2026}
\setcopyright{acmlicensed}\acmConference[CHI '26]{CHI}{Nov 5-9, 2026}{USA}
\acmBooktitle{CHI'24, Nov, 2026, USA}
\acmPrice{15.00}
\acmDOI{10.1145/3491102.43}
\acmISBN{978-1-4503-9157-3/22/04}

\begin{document}



\title[Bonik Somiti]{\textit{Bonik Somiti}: A Social-market Tool for Safe, Accountable, and Harmonious Informal E-Market Ecosystem in Bangladesh}

\author{ATM Mizanur Rahman}
\affiliation{
  \department{Computer Science}
  \institution{University of Illinois Urbana-Champaign}
  \city{Champaign}
  \state{Illinois}
  \country{USA}}
\email{amr12@illinois.edu}

\author{Sharifa Sultana}
\affiliation{
  \department{Computer Science}
  \institution{University of Illinois Urbana-Champaign}
  \city{Champaign}
  \state{Illinois}
  \country{USA}}
\email{sharifas@illinois.edu}

\renewcommand{\shortauthors}{Rahman et al.}

\begin{abstract}

People in informal e-markets often try to deal with fraud and financial harm by sharing posts, screenshots, and warnings in social media groups. However, buyers and sellers frequently face further problems because these reports are scattered, hard to verify, and rarely lead to resolution. We studied these issues through a survey with 124 participants and interviews with 36 buyers, sellers, and related stakeholders from Bangladesh and designed Bonik Somiti, a socio-technical system that supports structured reporting, admin-led mediation, and accountability in informal e-markets. Our evaluation with 32 participants revealed several challenges in managing fraud, resolving disputes, and building trust within existing informal practices and the assumptions behind them. Based on these findings, we further discuss how community-centered technologies can be designed to support safer and more accountable informal e-markets in the Global South.

\end{abstract}







\settopmatter{printfolios=true}

\maketitle

\input{1.intro}
\input{2.lit}
\input{3.phase-1}

\input{4.phase-2}

\input{5.phase-3-method}

\input{6.phase-3-feedback}

\input{7.discussion}
\input{8.lim-fw-con}



\bibliographystyle{ACM-Reference-Format}
\bibliography{sample-base}

\end{document}

%% file: 1.intro.tex
\section{Introduction}

Informal e-markets have become an important part of everyday economic life across the world. Nowadays, People use social media platforms not only to communicate but also to buy and sell goods, run small businesses, and earn income. Reports show that the social commerce industry was valued at around 492 billion US dollars in 2021 and is projected to grow to more than 1.2 trillion dollars in the coming years \cite{accenture_social_commerce_2022, mckinsey_social_commerce, shopify_global_ecommerce_stats}. At the same time, social media platforms have an unprecedented global reach, with more than 5.6 billion active users worldwide \cite{datareportal_digital_2025_social, datareportal_social_media_users}. Within this ecosystem, informal buying and selling activities take place at a massive scale. For example, \textit{Facebook} Marketplace alone reaches over one billion monthly visitors globally, while other platforms such as TikTok Shop generate hundreds of millions of dollars in sales during Black Friday in the United States \cite{meta_new_ways_to_shop_2021, engadget_facebook_marketplace_2021, businessinsider_tiktok_shop_2025}. These trends highlight the growing significance of informal e-markets within the global digital economy. 

Despite the global popularity informal e-markets are increasingly associated with fraud and financial harm. In the United States, consumer protection agencies report widespread financial losses from scams on social media platforms \cite{ftc_reported_losses_fraud_2025, ftc_top_scams_2025}. In the United Kingdom, authorities report hundreds of millions of pounds lost to online fraud each year \cite{uk-2025}. In Australia, scam monitoring agencies report widespread scam incidents and major financial losses due to scammers in informal e-market \cite{accenture_social_commerce_2022, scamwatch_social_media_scams}. International law warns that enforcement online scams are becoming more organized and increasingly rely on social media to reach victims at scale \cite{interpol_social_engineering_scams}. 
These risks are often more severe in many Global South contexts. For example, investigations and consumer protection reports have documented widespread scams on \textit{Facebook} Marketplace, including fake product listings, advance payment fraud, and courier-related scams that affect both buyers and sellers \cite{avast_facebook_marketplace_scams, abc_facebook_marketplace_scammers_2024, retailbanker_fb_marketplace_scams, news_com_au_facebook_phishing_scams}.  Government actions in the Global South report that many online commerce scams originate from social media marketplaces \cite{reuters_singapore_meta_anti_scam_2025}. News reports also show that thousands of \textit{Facebook} pages linked to marketplace scams have targeted users in countries such as India and Brazil \cite{toi_meta_takes_down_scams}. These trends reveal a growing gap between the expansion of informal e-markets in the Global South and the effectiveness of existing safeguards \cite{reuters_meta_ad_fraud_2025}.

Buyers and sellers in informal e-markets often rely on personal judgment, selective information sharing, and informal communication to manage fraud and risk \cite{mokhberi2025users}. However, weak verification, limited platform oversight, and low digital literacy make them highly vulnerable to scams and financial harm \cite{anjum2020drivers}. Prior research has mainly approached fraud as a transaction-level problem and focused on automated detection systems \cite{dong2021real}. While these methods show promising accuracy, they do not capture how fraud is experienced, negotiated, and managed within the everyday social practices of informal e-markets. As a result, we still lack a clear understanding of how risks are collectively handled and how design can better support these practices. We address this gap through a multi-phase study that examines informal e-market practices and explores design opportunities grounded in the lived realities of different stakeholders. We seek answers to the following research questions:
\begin{itemize}
    \item[] \textit{\textbf{RQ1:} What are the common and most prevalent types of fraudulent activities, risks, and harms experienced by buyers and sellers in informal e-markets in Bangladesh?}
    \item[] \textit{\textbf{RQ2:} What kinds of mechanisms and practices do buyers and sellers currently use to manage and respond to fraud and risk in informal e-markets, and what challenges often limit the effectiveness of these efforts?}
    \item[] \textit{\textbf{RQ3:} How can we design a socio-technical arrangement involving the broader informal e-market community to combat these challenges?}
\end{itemize}

To answer these questions we deployed a multi phase study. In the first phase of our study, we conducted an online survey (n = 124) and carried out in-depth interviews with 36 participants involved in informal e-markets in Bangladesh. Our participants included buyers, sellers, \textit{Facebook} fraud alert group admins, police officers, bank/fintech admins. Across these roles, participants reported several recurring types of risk and harm like reputation manipulation, product aspect authentication, transaction fraud etc.  After fraud incidents occurred, participants relied on informal coping strategies like public shaming, reporting in fraud alert groups etc to manage risk. However, many participants expressed frustration that these mechanisms rarely led to meaningful resolution.

Participants also described several challenges in combating risk. For example, unstructured dispute communication, unreliable fraud alert groups, identity switching to fraud people etc. We found participants of informal e-market operate in a high-risk environment with limited trust, weak accountability, and few effective pathways for resolution. The findings from the first phase of the study led to our second phase, where we aimed to design a system to better support informal e-market communities in managing fraud. Drawing from participants’ experiences, we identified four key design goals. G1 focuses on enabling structured reporting of informal e-market incidents. G2 aims to support administrative history keeping to help identify repeat and hidden offenders across changing identities. G3 seeks to facilitate admin-mediated conflict resolution between buyers and sellers. G4 focuses on enabling verified escalation of severe cases to formal authorities through community-validated evidence. With these goals in mind, we prototyped a socio-technical system to support fraud reporting, mediation, and accountability in informal e-markets. We evaluated the prototype through follow-up interview and focus group discussion (FGD) with 32 participants.

Our work makes a four-fold contribution to human–computer interaction (HCI), social computing, and research on informal digital marketplaces in the Global South. First, our findings provide an in-depth account of how buyers, sellers, and other stakeholders in Bangladesh experience scams and trust related breakdowns in informal e-markets, and how these risks affect everyday trading practices. Second, drawing on our empirical findings, we identify key challenges in existing risk-management practices and articulate a set of design goals that reflect the lived realities of informal e-market participants. Third, we design and prototype a socio-technical system and we evaluate this system with informal e-market stakeholders. Finally, through our engagement with participants, we surface broader socio-technical tensions surrounding trust, responsibility, and platform governance in informal e-markets, and we outline design implications for building safer and more accountable informal e-market infrastructures in the Global South.

%% file: 2.lit.tex
\section{Related Work}

In building our research questions and making arguments based on our findings, we draw on the literature of three broader themes: informal e-markets and everyday risk, vernacular trust and risk management in Global South informal markets, and informal design approaches that rely on community-centered socio-technical arrangements. In this section, we briefly review prior work across these themes and use it to situate our research questions on fraud, risk response, and community-based interventions in Bangladesh’s informal e-markets.

\subsection{Informal E-Markets and Everyday Risk}

Prior work shows that informal buying and selling is increasingly organized through mainstream platforms. Some of the most popular informal e-markets are social media-based. \textit{Facebook Marketplace} and Groups support local buying and selling across many regions, often involving individual and small-scale informal sellers \cite{fb1,fb2, meta_new_ways_to_shop_2021, engadget_facebook_marketplace_2021}. \textit{Instagram} serves as a visual storefront for informal commerce, especially for fashion, crafts, and food, with sales often happening through direct messaging and informal payment methods \cite{insta1, insta2}. Messaging apps like \textit{WhatsApp} and \textit{Telegram} are also used as direct sales channels, where people share product details, negotiate prices, and arrange delivery through personal conversations \cite{tandw1, tandw2, tandw3}. \textit{TikTok} has also expanded social commerce through live streams and direct interactions, which can appeal strongly to younger audiences \cite{tiktok1, tiktok2, businessinsider_tiktok_shop_2025}. Beyond these, many community and classifieds platforms in North America, such as \textit{Craigslist}, \textit{Nextdoor}, \textit{OfferUp}, \textit{eBay Classifieds}, \textit{VarageSale}, and \textit{5miles}, support local exchange through direct interactions \cite{lit1-minkus2014know, lit2-duh2002control, lit3-choksi2024under, lit4-masden2014tensions, lit7-kurwa2019building}. Similar patterns appear across regions, including \textit{Taobao} in China, \textit{Shopee} and \textit{Lazada} in Southeast Asia, \textit{Mercado Libre} in Latin America, and \textit{Jumia} in Africa \cite{china_tab, shop_and_laz, shop_and_laz2, mercado, jumia}. Together, this global landscape shows that informal e-markets are a mainstream way of doing commerce across many different platforms and regions. However, the ease of joining informal e-markets also makes everyday risk more common. Research in HCI and social computing has also shown how platform features such as advertising, post boosting, and paid content can shape visibility and affect trust and risk in these spaces \cite{voorveld2018engagement, shahbaznezhad2021role, eslami2016first}.  Reports from consumer protection agencies show that scams linked to social media platforms have caused widespread losses and have become more organized over time \cite{ftc_reported_losses_fraud_2025, ftc_top_scams_2025, interpol_social_engineering_scams}. Taken together, this literature suggests a growing gap between the rapid expansion of informal e-markets and the ability of existing systems to prevent fraud and protect everyday participants. This body of work motivates a closer examination of what kinds of fraud, risks, and harms are most prevalent in informal e-markets, and why these harms persist across platforms and regions (RQ1). 

\subsection{Global South Informal E-Markets and Vernacular Risk Management}

Informal e-markets in the Global South are often sustained through vernacular practices, where people adapt digital technologies to local norms, infrastructures, and everyday constraints \cite{chandra2017bazaar, pal2018digital, bhattacharya2019ict}. Prior work on South Asian marketplaces shows that trust is built through embodied and socially grounded practices such as bargaining, product testing, and repeated interactions \cite{chandra2017bazaar, pal2018digital}. 
Instead of adopting rigid formal infrastructures, market actors often maintain informal coordination and community control by using whatever tools fit their context, such as intercoms, messaging apps, or small ad-hoc networks \cite{pal2018digital}. Joshi et al. describe this as a form of translation, where textile sellers in Surat reshape structured e-market workflows into flexible WhatsApp exchanges that better match local social norms and collaboration styles \cite{joshi2025reselling}. Related work in Kenya also shows how Nairobi youth combine \textit{Facebook} and other ICT practices to support everyday economic goals  \cite{wyche2013hustling}. These studies show that informal e-market actors manage uncertainty by assembling their own tool ecologies, rather than relying on a single platform to provide safety, structure, or accountability.

However, this flexibility can create challenges for effective fraud and risk response. Bhattacharya cautions against assuming that informality is simply a temporary lack of modernization. Instead, informality can be a rational response to structural exclusion, where actors prioritize stability, privacy, and control over scale \cite{bhattacharya2019ict}. Prior research also emphasizes that support systems are more effective when they are participatory and context-aware \cite{rumanyika2022design}. Work on infrastructural precarity similarly highlights that technology solutions can fail when they assume stable access, stable rules, or stable institutional support that does not exist in many Global South settings \cite{ebrahim2024barriers}. Research on transnational communities shows how people develop informal digital solutions to overcome financial exclusions, demonstrating how coping strategies emerge when formal infrastructures are unavailable \cite{rohanifar2021money, rohanifar2022kabootar}.

In Bangladesh, informal e-market activity reflects this same pattern of vernacular adaptation and trust work. Sellers benefit from interactional intimacy on social media, but they also face trust and logistical challenges that push them to develop layered strategies \cite{clemes2014empirical, yahaya2021mediating}. Studies in Thailand similarly show that personalization can shape buyer trust \cite{driediger2019online}. Research in Dhaka further confirms that social validation signals, such as engagement and visible endorsement, strongly influence consumer decisions \cite{enam2024online, suhan2015acceptance}. These findings suggest that buyers and sellers in Bangladesh often manage risk through social connections, visible trust cues, and improvised verification practices, but these practices can also be fragile, inconsistent, and easy to manipulate. This body of literature motivates the need to examine the specific mechanisms that buyers and sellers currently use to respond to fraud and risk in informal e-markets, as well as the practical challenges that limit these efforts. It also suggests that effective interventions should build on vernacular practices and community-based trust work (RQ2, RQ3).

\subsection{Informal Design and Community-Centered Socio-Technical Arrangements}

Prior work in HCI, CSCW, and ICTD shows the importance of informal design. In many Global South settings, informality is a practical way that communities organize work, care, and coordination when formal systems are inaccessible \cite{rangaswamy2011cutting, sambasivan2010intermediated, bhattacharya2019ict}. People often sustain everyday activities through shared access, creative workarounds, and locally meaningful routines \cite{rangaswamy2011cutting}. A consistent theme across informal design scholarship is that communities develop their own trust and coordination infrastructures when formal services fail them. For example, when formal FinTech systems exclude politically marginalized immigrants, people turn to informal and risky cross-border money transfers \cite{rohanifar2021money}. Kabootar demonstrates how peer-to-peer coordination and social trust can serve as key design resources in such contexts \cite{rohanifar2022kabootar}. Related marketplace studies similarly show that communities rely on informal social infrastructures to manage uncertainty \cite{chandra2019rumors}. In Bangladesh, Protibadi enables collective visibility around public sexual harassment through shared narratives \cite{ahmed2014protibadi}. ShishuShurokkha demonstrates the value of community-centered and bystander-based reporting in conservative contexts \cite{sultana2022shishushurokkha}. Unmochon further shows how informal justice practices support evidence sharing within trusted networks \cite{sultana2021unmochon}. Together, these studies highlight both the promise and complexity of community-based harm reporting when formal systems fail.

Prior informal design research shows that systems work better in informal settings when they are grounded in local knowledge, lived experience, and meaningful participation. Studies in health and public services show that people often rely on existing social networks and community relationships to access information \cite{kumar2015mobile, ramachandran2010mobile, molapo2016designing, ismail2018bridging}. Other work highlights the role of intermediaries who help others participate despite barriers, while also reminding us that such roles can introduce new power dynamics and responsibilities \cite{sambasivan2010intermediated, sambasivan2011designing}. Recent studies caution that community participation should be designed with care, so that support does not turn into surveillance or added burden \cite{cserban2025less, petterson2023playing}. Together, this body of work suggests that community-centered designs should gently balance participation, responsibility, and protection in informal settings.

Taken together, Informal design scholarship suggests that when formal safeguards are weak, communities often rely on social trust, intermediaries, collective sensemaking, and locally grounded accountability practices to manage risk and harm \cite{rohanifar2022kabootar, chandra2019rumors, sambasivan2010intermediated}. At the same time, research on community-based reporting and justice-oriented designs shows that informal accountability can be fragile and can introduce new challenges, such as privacy risks, unequal burdens, and contested authority \cite{sultana2021unmochon, petterson2023playing}. However, we still have limited understanding of how these dynamics play out in informal e-markets in Bangladesh, and how a community-centered socio-technical arrangement can distribute responsibilities and strengthen accountability without undermining the flexibility of informal e-market (RQ3).

%% file: 3.phase-1.tex
\section{Understanding Fraud and Deceptive Practices in Bangladeshi Informal E-Markets}

We conducted an anonymous survey with 124 participants to investigate the nature of risks and fraudulent activities in Bangladeshi informal e-markets, the strategies participants use to mitigate these risks, and the challenges they face in combating fraud despite employing such strategies. We then conducted in-depth interviews with 36 participants to develop a deeper understanding of these experiences from multiple perspectives and to identify potential design opportunities. All researchers involved in this project were born and raised in Bangladesh and are native Bengali speakers. These criteria helped the researchers engage with and understand the participants better. This study was approved by the ethics review committee of the authors’ institutions. This section describes the methods and related findings from this phase of the work in detail.

\subsection{Methods}

We conducted both an online survey and an interview study to understand fraud and deceptive activities in Bangladeshi informal e-markets, as well as the strategies participants use to resist these practices and the challenges involved. The anonymous online survey enabled participants to report a wide range of experiences with risk and fraud across informal e-markets. In contrast, the interviews allowed us to examine specific incidents in depth and to understand participants’ reasoning, responses, and constraints in navigating deceptive activities. We describe both of these methods in further detail below.

\subsubsection{Online Survey.}

We hosted our anonymous online survey on Google Survey and circulated it online through multiple channels, including emails to the authors’ personal networks, \textit{Facebook} direct messages to contacts, and buyer–seller \textit{Facebook} groups commonly used for informal e-market transactions. Participants accessed the survey by clicking a link shared via email or \textit{Facebook} posts. Participant consent was obtained within the survey itself; after reading the consent information, participants were able to proceed only after agreeing to participate. The survey consisted of three sections tailored to different stakeholder roles in informal e-markets: buyers, sellers, and fraud detection group admins. For buyers and sellers, the survey asked about the types of fraudulent activities they encountered in informal e-markets, the strategies they used to mitigate such risks, their reporting mechanism when fraud occurred and other related information. For fraud group admins, the survey focused on the types of posts they encountered in fraud-alert groups, challenges they faced in moderating, verifying, or responding to such posts and other relevant information. 

Most questions were structured as check-box or multiple-choice items; however, the survey also included several optional open-ended text boxes that allowed participants to elaborate on their experiences. No personally identifiable information (e.g., name, precise location) was collected from participants. We did not offer any compensation for participation. Completing the survey typically took approximately 10–15 minutes. Participants were free to exit the survey at any time by closing their browser window without any consequences. In total, 124 participants completed the survey. Table ~\ref{pre-demo} summarizes participants’ age range, gender, role in informal e-markets , the platforms they used, and the length of time they had been engaged in informal e-market activities.

\subsubsection{Interviews.}

To gain a deeper understanding of fraud and deceptive practices in Bangladeshi informal e-markets, we conducted 36 semi-structured interviews with participants involved in different roles. Among them, 9 participants were both buyers and sellers, 11 were buyers only, 8 were sellers only, and 4 were group admins of fraud alert groups, 2 were police officers and 2 were Bank/Fintech workers. We recruited participants through snowball sampling \cite{goodman1961snowball}. Each interview lasted between 30 and 60 minutes. Interviews were conducted either in person or via Zoom, depending on participant preference. All interviews were conducted in Bengali, the native language of both the participants and the researchers. In some cases, we used a translated version of the interview protocol to ensure clarity and participant comfort. At the beginning of each interview, we explained the purpose of the study and emphasized that participation was voluntary. We obtained oral consent from all participants. Participants were informed that their identities would remain anonymous and that they could withdraw from the study at any time without any consequences. During the interviews, we asked participants about their experiences with fraud and deceptive activities in informal e-markets, including incidents they encountered while buying or selling goods. We also discussed how they identified risky situations, the steps they took to respond to fraud, and where they sought help or reported incidents. For group adminis, we focused on their experiences managing fraud-related posts, verifying claims, and handling disputes within community groups.

\subsubsection{Data Collection and Analysis.}

All survey responses were initially collected through Google Survey. The responses were anonymized by default and retrieved in comma-separated values (CSV) format. We then stored the data in a secure storage space accessible only to the research team. We used open-source statistical tools in Python to analyze the survey data. In addition to descriptive analysis, we qualitatively examined open-ended responses to better understand participants’ experiences with fraud, risk, and deceptive practices in informal e-markets.

The interview study generated a total of approximately 16 hours of audio recordings. We transferred all recordings to a secure computer owned by the researchers. The interviews were then transcribed and translated into English. The interview transcripts resulted in approximately 120 pages of documented data. All identifying information was removed before conducting open coding and thematic analysis on them \cite{boyatzis1998transforming, strauss1990basics}. Two authors independently read through the transcripts and allowed code to develop. Later they shared their codes with each other. A total of 32 codes developed initially including fraud, transaction, resolution, no support, harassment, risk etc. After a few iterations, we clustered related codes into themes and drew our design assumptions on them.

\begin{table}[t]
\centering
\renewcommand{\arraystretch}{1.3}
\setlength{\tabcolsep}{6pt}

\begin{tabular}{|p{1.5cm}|p{3.0cm}|p{4.0cm}|p{4.0cm}|}
\hline

\multicolumn{4}{|c|}{\textbf{Total Number of Survey Responses: 124 (Female: 76, Male: 48)}} \\ \hline

\textbf{Age Range (in Years)} &
\textbf{Participant Roles} &
\textbf{Years in Informal E-Market} &
\textbf{Platform Used by Participant} \\ \hline

18--25: 51  \newline
26--35: 62  \newline
36--45: 8  \newline
46+: 3  &
Buyer: 51  \newline
Seller: 24  \newline
Buyer+Seller: 36  \newline
Fraud Group Admin: 13  &
Less than 6 months: 8  \newline
6 months - 1 year: 26  \newline
1 - 3 years: 48  \newline
More than 3 years: 42 &
\textit{Facebook} Only: 102  \newline
\textit{Facebook+Instagram}: 12  \newline
\textit{Facebook+WhatsApp}: 6  \newline
All Three Above: 4 \\ \hline

\multicolumn{4}{|c|}{\textbf{Total Number of Interview Participants: 36 (Female: 19, Male: 17)}} \\ \hline

\textbf{Age Range (in Years)} &
\textbf{Participant Roles} &
\textbf{Years in Informal E-Market} &
\textbf{Platform Used by Participant} \\ \hline

18--25: 12  \newline
26--35: 19  \newline
36--45: 3  \newline
46+: 2  &
Buyer: 11  \newline
Seller: 8  \newline
Buyer+Seller: 9  \newline
Fraud Group Admin: 4 \newline 
Police Officer: 2  \newline
Bank/Fintech Worker: 2  &
Less than 6 months: 3  \newline
6 months - 1 year: 6  \newline
1 - 3 years: 10  \newline
More than 3 years: 13 &
\textit{Facebook Only}: 24  \newline
\textit{Facebook+Instagram}: 5  \newline
\textit{Facebook+WhatsApp}: 2  \newline
All Three Above: 1 \\ \hline

\end{tabular}

\caption{Demographic Characteristics of the Pre-design Survey and Interview Participants}
\label{pre-demo}
\end{table}

\subsection{Unpacking Aspects of Fraud and Deceptive Practices in Bangladeshi Informal E-Markets}

The findings from the survey and interviews help us understand the key actors involved in informal e-market conflicts, the types of fraud and the risks they experience. Our findings also reveal how participants manage these risks and the challenges they face in doing so. This subsection highlights the key findings. 

\subsubsection{Actors in Informal E-market Conflicts}

This subsection introduces the three main actors involved in conflicts around fraud and deception in Bangladeshi informal e-markets: buyers, sellers, and fraud alert group administrators. We describe who these actors are, how they participate in informal e-market transactions, and how frequently they encounter or deal with fraud and other related challanges. 

\textbf{(a) Buyers.} In Bangladeshi informal e-markets, buyers are people who purchase products directly from sellers through social media platforms. From our survey, 87 participants identified themselves as buyers. We asked buyers about the amount of money they had lost due to scams while purchasing items through informal e-markets. Reported losses ranged from BDT. 0 to BDT. 60,000 (USD. 500), with an average loss of approximately BDT. 3,500 (USD. 30) per buyer. We also asked buyers to reflect on how often they encountered fraud during transactions. On average, buyers reported that 2.7 out of every 10 transactions involved some form of fraud or deceptive behavior. These findings indicate that fraudulent activities are a recurring part of everyday buying experiences in Bangladeshi informal e-markets. 

\textbf{(b) Sellers.} In Bangladeshi informal e-markets, sellers are individuals who offer products or services directly to customers through social media platforms. From our survey, 60 participants identified themselves as sellers. Sellers also reported experiencing scams while selling products. Reported monetary losses ranged from BDT. 0 to BDT. 85,000 (USD. 700), with an average loss of approximately BDT. 6,000 (USD. 50) per seller. We also asked sellers about how often they encountered deceptive behavior during transactions. On average, sellers reported that 3.2 out of every 10 transactions involved some form of fraud or deceptive practice. These findings suggest that similar to buyers, sellers also face financial risk in informal e-markets.

\begin{figure}[t]
    \centering
    \includegraphics[width=0.9\linewidth]{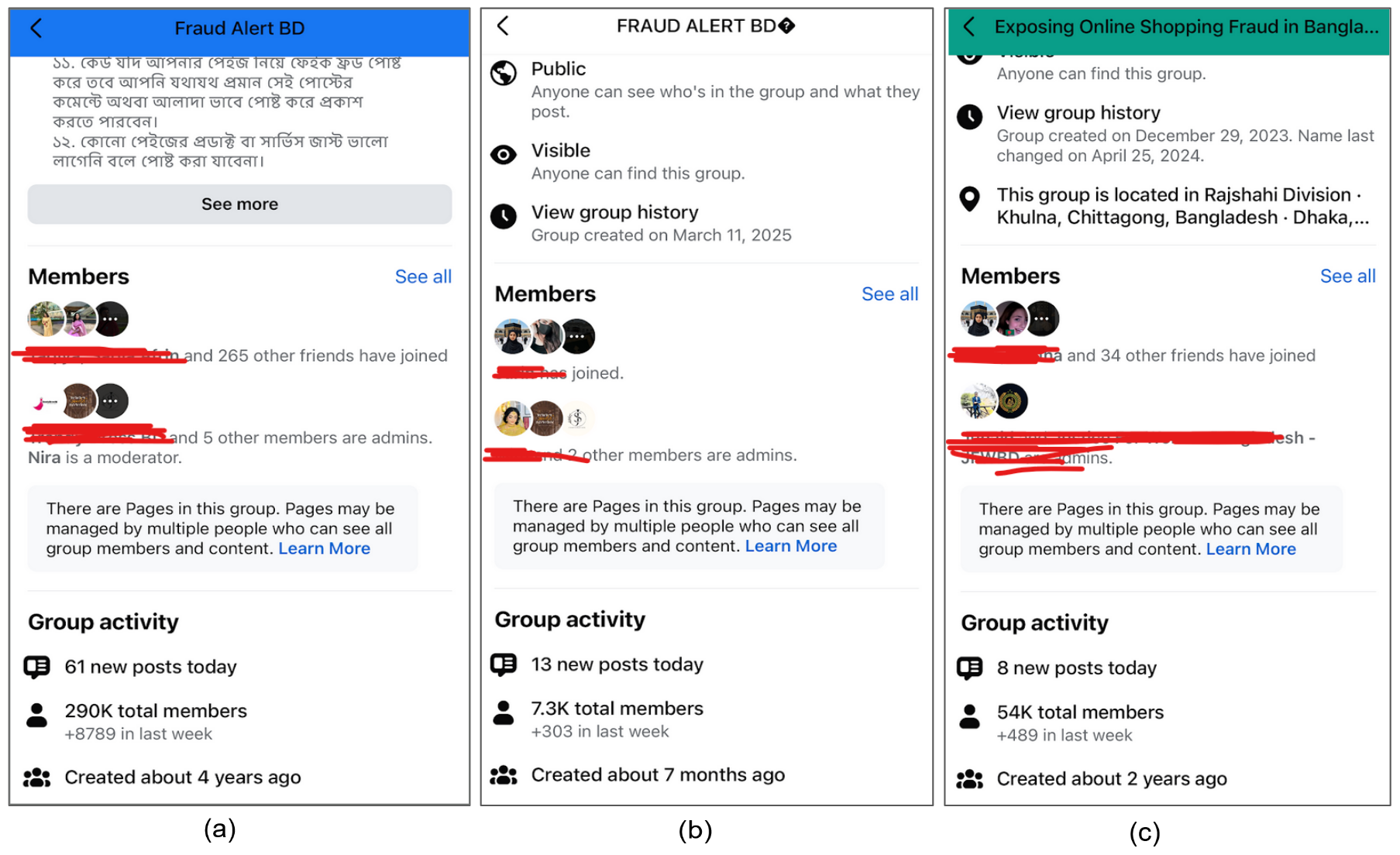}
    \caption{Screenshots of three \textit{Facebook} fraud alert groups (shared by the group admins) in Bangladesh with different sizes and activity levels. The left group (a) shows very high daily activity (61 new posts per day) and a large number of members (around 290,000), indicating frequent fraud incidents and widespread public concern. The middle (b) and right (c) groups also show regular daily posts and tens of thousands of members, suggesting that many users rely on such groups to report scams and stay aware. These groups highlight the severity and scale of fraud in informal e-markets.}
    \label{fig:fraud-groups}
\end{figure}

\textbf{(c) Fraud Group Admins.} 

In Bangladeshi informal e-markets, fraud group administrators play an important role in managing community-led responses to scams. On \textit{Facebook}, several fraud alert groups exist where buyers and sellers share incidents they have faced. The goal of these groups is to warn others and reduce future fraud. Many of these groups were created by experienced sellers who had repeatedly faced scams and wanted to protect others in the market. P13 explained how these groups first emerged as small and private spaces among sellers

\begin{quote}
\textit{``As a seller, I faced various types of scams. Often, buyers did not pay after placing an order. Some would block me after receiving the product, while others would receive the product and then claim that they never got it. Scams were happening everywhere. Because of this, some fellow sellers and I first created a Messenger group to warn each other. Whenever any of us became a victim of a scam, we shared the buyer’s \textit{Facebook} ID in the group so that others could stay aware. As more sellers learned about this strategy, they wanted to join our Messenger group. We realized that since so many people were interested creating a \textit{Facebook} group would with sellers would be a better approach." \textbf{(P13)}}
\end{quote}

Over time, some of these groups expanded beyond sellers and became public. Another admin described this shift

\begin{quote}
\textit{``At first, our group was only for sellers. Later, we realized that there were many fraudulent sellers in the market who were also scamming buyers. This affected me personally because I sell authentic makeup products that I import from other countries, but my sales were decreasing due to fraudulent sellers who sold duplicate products. Customers often mentioned these pages to me when raising concerns about product authenticity. Because of this, we decided to make the group public, where both buyers and sellers could post, share warnings, and alert others about fraudulent activities." \textbf{(P19)}}
\end{quote}

Fraud alert group admins play a central role in keeping these groups active. Based on our interviews, admins receive a high number of requests every day, often between 10 and 100 posts, depending on the group size and activity level. While sharing our survey with fraud group admins, we found at least 20+ fraud alert groups, ranging from around 5,000 to 320,000 members (See figure ~\ref{fig:fraud-groups}). This scale shows that fraud alert groups and their admins have become an important part of how people manage fraud risks in Bangladeshi informal e-markets.

\subsubsection{Types of Scams and Associated Risks in Bangladeshi Informal E-Market}

This subsection describes the different types of risks and scams experienced by buyers and sellers in Bangladeshi informal e-markets. 

\textbf{(a) Transaction Fraud Experienced by Both Buyers and Sellers}

From our survey, 103 participants reported experiencing transaction fraud at least once in informal e-markets. Sellers described several forms of fraud: 32 reported buyers who failed to pay or canceled orders after products were already shipped, and 28 reported cases where buyers received products but did not pay or blocked the seller. Buyers reported similar problems. Fifty-two buyers said they paid in full but never received the product, while 18 reported partial advance payment fraud, where they paid 20–50\% in advance but the seller did not deliver the product.

Beyond these common forms of transaction fraud, our interviews revealed a newer and more concerning practice involving fake payment receipts generated using AI tools. Four sellers shared that they had personally experienced fraud involving fake digital payment receipts. In addition, three fraud-monitoring group administrators reported seeing multiple posts in their groups where sellers described similar incidents involving AI-generated receipts. One seller P4 explained how this happened to him:

\begin{quote}
\textit{``After the buyer placed the order, he shared a screenshot of a payment receipt with me, and I also received a transaction SMS on my bKash number. Because of this, I assumed the payment had been completed and delivered the order. A few days later, while calculating my sales, I noticed that some money was missing. I contacted a bKash agent and shared the receipt screenshot and the SMS. They confirmed that no such payment had been made and that the receipt was fake, likely generated using LLM. As a result, I lost the money, and they informed me that they were not responsible for this loss. I do not know exactly how this was done, but recently I have heard about similar cases from other sellers as well." \textbf{(P4)}}
\end{quote}
 
These findings show that transaction fraud in informal e-markets takes multiple forms and affects both sides of the transaction. The absence of verified payment confirmation, reliable dispute resolution mechanisms, and trusted intermediaries makes it easy for fraud to occur. 

\textbf{(b) Reputation Manipulation Experienced by Sellers}

From our survey, 36 sellers reported experiencing different forms of reputation manipulation by buyers. These practices often involve false reporting, coordinated attacks, and misuse of platform features, which directly affect sellers’ visibility, reach, and trustworthiness in the market. In many cases, these reports were made out of anger, personal grudges, or without any clear reason. 

Thirteen sellers shared that reputation manipulation also happens due to misunderstandings between buyers and sellers. Sellers explained that even small communication gaps can quickly escalate into public reputation damage. In such cases, buyers respond by posting negative comments or reviews instead of trying to resolve the issue privately. One seller shared an experience where a misunderstanding led to a coordinated reputation attack:

\begin{quote}
\textit{``When I was selling to a buyer, I told them that my products come from outside Dhaka. However, the buyer misunderstood this and thought that I import the products from abroad. Because of this misunderstanding, the buyer started abusing me and arguing with me. The buyer and his friends did mass reporting to my page and posted offensive comments, left fake negative reviews, and wrote false comments about my products. Due to this my reach of page reduced a lot and my reputation got damaged." \textbf{(P12)}}
\end{quote}
 
These findings show that seller reputation in informal e-markets is highly fragile and can be easily manipulated by buyers. The lack of reliable moderation and absence of fair review or appeal mechanisms make it difficult for sellers to protect themselves against false and coordinated attacks.




\textbf{(c) Product Authenticity and Misrepresentation Experienced by Buyers}

From our survey, 52 buyers reported facing issues related to product authenticity. Buyers explained that the products they received often did not match what was shown in images or descriptions. Thirty-eight buyers reported that sellers used edited images, AR filters, or image overlays to hide product defects or falsely display brand logos. In addition, 28 buyers mentioned noticing changes in product quality after delivery, including lower fabric quality in clothing, differences in color, and overall degradation in material compared to what was shown online.


 
From our interviews, seven buyers reported experiencing deception through AI-generated content. Buyers explained that sellers increasingly use AI or LLM to generate visually attractive images or videos. P3 described such an experience:

\begin{quote}
\textit{``I saw an Islamic calligraphy frame online. The image looked very premium, with perfect lighting and design so I decided to place an order. However, when I received the product, I saw that the finishing, material, and overall look were completely different from what was shown in the image. I messaged the seller and told them that the product did not look like the image they advertised. They replied that they could not provide something that was 100\% the same and said that they use attractive images on their page to draw more attention and increase engagement. I told them that I was willing to pay more if they could give me the exact product shown in the image. Later, they admitted that the image was actually generated using AI and that they do not have artists who can create calligraphy like that." \textbf{(P3)}}
\end{quote}
 
These findings show that product authentication in informal e-markets is heavily manipulated. The lack of standardized product descriptions, verification mechanisms, and accountability allows sellers to fraud without consequence. 

\textbf{(d) Product Delivery Complications}

From our survey, 62 buyers reported facing delivery-related problems. Forty-one buyers stated that their products were damaged during delivery by courier companies. In these cases, buyers explained that neither the seller nor the courier company accepted responsibility for the damage. Forty-five buyers reported that when the delivered product did not meet their expectations, sellers refused to accept returns, even when they had initially agreed to a return policy. In addition, 17 buyers stated that even after successfully returning the product, they did not receive any refund from the seller.

Delayed delivery was another major concern for buyers. Thirty-two buyers reported being severely affected by late product delivery caused by either the seller or the courier company. Buyers shared examples such as birthday cakes arriving after the birthday, makeup products delivered after wedding functions, food arriving after events had ended, medicines for elderly family members arriving late, items ordered before travel arriving after flights, and anniversary flowers being delivered after midnight the next day. From the interview P15 described how a delayed delivery affected her on an important life event:

\begin{quote}
\textit{``I ordered a customized yellow sari (a traditional dress) for my holud (a special pre-wedding ceremony) one week in advance. The seller initially told me that I would receive the sari one day before the ceremony. However, when I contacted them again the day before, they said I would receive it on the day of the ceremony. However, the sari arrived one day later, but by then it was useless. On the day of my gaye holud, my brother had to quickly buy a sari from a shopping mall, and I wore that instead." \textbf{(P15)}}
\end{quote}

Overall, our findings show that delivery failures in informal e-markets extend beyond logistical inconvenience and can have lasting personal, financial, and health-related impacts on buyers’ lives.



 

\subsubsection{Existing Way of Managing Risks}

This subsection focuses on the strategies that buyers and sellers currently use to manage risks in informal e-markets. Participants rely on a range of informal practices which are described below. 

\textbf{(a) Public Shaming}

From our survey, 36 participants reported using public shaming as a way to respond to deceptive behavior by sellers. Buyers explained that one common approach is to comment directly on the seller’s page, describing what happened to them and warning others about the fraudulent activity. In more severe cases, 11 buyers reported posting about the incident on their own public \textit{Facebook} profiles. By making fraud visible in public spaces, buyers attempt to create informal accountability and protect others fram the scams in informal e-market.

\textbf{(b) Reporting in Fraud Alert Groups}

From our survey, 65 participants mentioned that reporting in fraud alert groups is their primary response after experiencing a fraudulent incident. Participants explained that after facing fraud, they usually collect evidence such as screenshots of messages, payment details, and seller or buyer profiles. They then write a short description of the incident and post it in \textit{Facebook}-based fraud groups. The main goal of sharing these posts is to warn others and prevent similar scams from happening again. In the absence of official reporting, these fraud groups act as informal spaces where users exchange information and collectively respond to risks in informal e-markets.

\textbf{(c) Seeking Formal Remedies}

From our survey, Eight participants mentioned that they reported incidents to the police, hoping to recover their money or take action against the fraudulent party. Twelve participants reported contacting banks or fintech services to report fraudulent transactions or seek assistance with payment-related issues. In addition, four participants reported reaching out to courier companies when delivery-related problems occurred.
Although these formal remedies are seen as potential solutions, many reported that these processes do not result in meaningful resolution.

\textbf{(d) Private Warnings Within Trusted Networks}

From our survey, 24 participants mentioned that they respond to fraudulent experiences by privately informing close friends or family members through direct messages. Participants explained that instead of posting publicly, they prefer to share details of the incident with people they know and trust. This approach is often chosen by participants who feel uncomfortable with public shaming or who want to avoid further conflict.
Private warnings serve as a low-risk and personal strategy for managing harm.

\subsubsection{Challenges in Combating Risks}

This subsection examines why risks in informal e-markets continue to persist despite existing mitigation strategies. Our findings show that fragmentation, lack of accountability, unclear complaint mechanisms, and identity fluidity limit the effectiveness of current responses.

\textbf{(a) Challenges in Fraud Alert Groups}

Our findings show that although fraud alert groups are commonly used as a risk mitigation strategy, they are not very effective. From our survey, 43 participants said there are too many fraud alert groups on \textit{Facebook} with similar names, which makes it confusing to know which group to trust. From our interviews, 17 participants also reported problems with searching within these groups. Participants reported that searching for specific names, page titles, or phone numbers often does not return relevant results. This makes it very difficult to verify whether a buyer or seller has been reported multiple times or has a history of fraudulent behavior. P24 described shared one expereince:

\begin{quote}
\textit{``Before buying anything online, I usually search alert groups to check whether a page has any negative reports or complaints. In this case, I searched the name of a three-piece clothing page in three or four fraud groups and did not find any complaints. So I placed the order. However, when I received the product, I saw that the quality was very poor. As soon as I washed the fabric, all the colors faded. When I tried to return the item, they refused, and I lost the full amount. Later, I shared the issue with a friend. When my friend searched the page name in the fraud groups, the complaints appeared immediately. I then searched again and found them as well. If those reports had appeared when I searched the first time, I would not have suffered this loss." \textbf{(P24)}}
\end{quote}
 
Participants also questioned the authenticity and reliability of posts shared in fraud alert groups. From our survey, 48 participants stated that they do not fully trust posts in these groups because they believe false or fake accusations are common. At the same time, fraud group administrators acknowledged the high frequency of fake posts and their limited control over posts. From our interviews, seven admins explained that because these groups operate on \textit{Facebook}, they do not have the power to store or preserve posts. As one admin explained:

\begin{quote}
\textit{``Every day, we receive many fake posts, and we identify those posts and delete them. In some situations, we approve a post, but later the person who made the accusation deletes it. After that, other people ask us why we deleted the post. It becomes difficult to explain that we did not delete it and that the accuser removed it on their own." \textbf{(P14)}}
\end{quote}
 

 
These findings show that while fraud alert groups are intended to support risk mitigation in informal e-markets, they are constrained by fragmentation, poor search functions, questionable authenticity, and inconsistent moderation. The lack of standardized verification and accountability within these groups makes it difficult for participants to confidently identify real fraud cases.

\textbf{(b) Identity Fluidity in Informal E-Markets}

From our survey, 42 buyers reported that frequent changes in identity make it very difficult to avoid scams or hold anyone accountable. Participants explained that in informal e-markets, sellers can easily change their page name, personal profile, phone number, or fintech number and continue operating under a new identity. This concern was also strongly echoed by fraud group administrators. All fraud group admins we interviewed echoed this concern. One admin described this pattern:

\begin{quote}
\textit{``It is very difficult to track scammers using their page names. A few months ago, I noticed several complaints against a \textit{Facebook} page selling women’s handbags and purses, mostly about taking advance payments and not delivering products. After seeing multiple posts, I shared a warning as an admin in a seller-buyer group and asked people to stay alert. A few months later, I started seeing similar complaints again, this time about women’s purse bags from a different page. When I checked, I found the same product photos and identical captions, but a different page name, admin name, and contact number. People do this a lot, and I do not know how many times I can actually fight this." \textbf{(P13)}}
\end{quote}
 
These findings show that identity fluidity allows fraudulent actors to repeatedly re-enter informal e-markets without facing long-term consequences. The ease of creating new pages and changing contact details undermines the effectiveness of fraud reports, public warnings, and group-based monitoring.

\textbf{(c) Absence of Third-Party Mediation}

From our interviews, seven participants emphasized the absence of any neutral party to handle disputes or facilitate communication between buyers and sellers. They explained that when a dispute occurs, there is no trusted person who can listen to both sides and help resolve the issue fairly. As a result, misunderstandings are hard to clarify and reaching mutual agreement becomes challenging. One seller described how this situation affects her

\begin{quote}
\textit{``When selling products, sometimes mistakes happen. However, buyers often do not try to understand this. Instead, they immediately post negative complaints using our page name, say harmful things, and publicly shame us. This damages our reputation. In many cases, there is no way to explain to them that the issue was a genuine mistake from our side and that we are willing to fix it." \textbf{(P35)}}
\end{quote}
 
In addition, five participants mentioned that the lack of a trusted middleman makes them hesitant to buy or sell in informal e-markets. Participants reported that direct communication sometimes leads to threats or harmful behavior, which discourages them from engaging further.




\textbf{(d) Absence of Effective Complaint Mechanisms}

From our survey, 73 participants reported that when scams occurred, they did not know how, where, or to whom they could complain. In total 30 participants mentioned confusion about whether police complaints are applicable to informal e-market fraud. In addition, 23 participants explicitly stated that they avoided complaining because they did not want to become involved in legal processes. From our interviews, three participants shared that even when they attempted to file police complaints, they did not receive supportive responses. From our interviews with Police officers we received similar responses. As One officer P29 said

\begin{quote}
\textit{``There is no formal policy for handling fraud in informal e-markets in Bangladesh. If a scam happens, we are not officially bound to work on these cases. Sometimes, out of human values and concern for people’s safety, we try to help. But the problem is, if we take one case, many others will come to us with similar complaints. We do not have enough time or manpower for that. Verifying every single unverified case from informal online markets is extremely difficult and would be a huge waste of time for us." \textbf{(P29)}}
\end{quote}

Our interviews further revealed that participants who attempted to seek help from financial institutions were largely unsuccessful. Twelve participants reported contacting banks or fintech services such as bKash and Nagad after experiencing fraud. However, they explained that these institutions were unable to provide meaningful support. One participant described her experience with a bank:

\begin{quote}
\textit{``After making the payment using my credit card, I realized that the entire transaction was a scam. I called the bank and explained everything in detail. I shared all the information with them, including proof of the scam and the receiver’s details, and asked whether there was any way to get my money back. However, they told me that banks in Bangladesh do not have any policy that allows a payment to be reversed once it has been completed. As a result, they said they could not help me recover the money." \textbf{(P16)}}
\end{quote}

During our interviews with bank and fintech workers, we asked about their policies for handling such fraud cases. One bank worker P31 explained:

\begin{quote}
\textit{``Banks, credit card companies, and fintech services generally do not have any policy to reverse a transaction or return money once the payment is completed. It becomes the responsibility of users because they made the transaction themselves, so we are not allowed to reverse it afterward. However, in severe cases, we may be able to freeze the recipient’s account temporarily." \textbf{(P31)}}
\end{quote}


 
These findings show that the lack of accessible, trusted, and effective complaint mechanisms prevents participants from seeking justice after fraud occurs. Unclear legal frameworks, limited support from financial institutions, and discouraging responses from law enforcement remains as a major challenge in informal e-market.

\subsection{Summary and What Is Missing}

In summary, we found that buyers and sellers in Bangladeshi informal e-markets frequently face fraud, delivery problems, and misleading product listings. When such incidents occur, there is no formal way to report or document them. Instead, people rely on posting screenshots in \textit{Facebook} fraud alert groups, warning others privately, or publicly accusing alleged offenders. These practices are scattered and inconsistent, and many reports fade away over time. Community admins explained that this makes moderation very difficult. They lack any system to keep records of past incidents, which allows repeat offenders to change names or identities and continue operating without detection. At the same time, there is no neutral space where buyers and sellers can resolve misunderstandings through mediated discussion. Participants also noted that some cases are too severe to be handled within the community and require help from formal authorities. However, individual reports are often ignored because evidence is hard to verify and authorities are reluctant to act. Thus, we found that the key challenges in informal e-markets include the absence of structured reporting, the lack of administrative history to identify repeat or hidden offenders, limited support for mediated dispute resolution, and the absence of verified pathways for escalation to formal authorities. We translated these challenges into our design goals, which guided the design of our system, as detailed in the next section.

%% file: 4.phase-2.tex


\section{Design of Bonik Somiti}

Drawing on the findings from the survey responses and interviews, we designed a tool, namely Bonik Somiti. Bonik Somiti is a bengali word which represents traditional traders’ association in Bangladeshi physical markets that mediates disputes between buyers and sellers through community-based negotiation (salishi) and informal governance. While discussing the idea of a hypothetical tool that would help address fraud, disputes, and accountability in informal e-markets, one of the participants named it Bonik Somiti. We borrowed the name from them. This section describes our design goals and components of the application.

\subsection{Design Goals}

Building on the findings from Phase 1, we identified several structural challenges in informal e-markets. Participants repeatedly highlighted the absence of any systematic reporting mechanism, lack of historical records to track repeat offenders, unresolved conflicts escalating into public shaming, and limited pathways to escalate severe cases to formal authorities. Therefore, we set our objective to design a community-centered system that supports reporting, verification, mediation, and escalation. Based on these findings, we determine four design goals.

\textbf{G1: Enable Structured Reporting of Informal E-Market Incidents.}

Participants described that for most negative occurrences in informal e-markets there was no formal or systematic way to report or file a case. This urges us to design a reporting mechanism where both buyers and sellers can formally report incidents through the system. While reporting, users can provide incident details, supporting screenshots, and relevant contextual information, which are then forwarded to community admins for verification.

\textbf{G2: Support Administrative History Keeping for Identifying Repeat and Hidden Offenders.}

Participants explained that fraudulent sellers or buyers could easily change \textit{Facebook} page names, contact information, or identities and continue operating without detection. This scenario urges us to design a history-keeping mechanism accessible only to community admins, where records are organized based on identifiers such as phone numbers, page names, page URLs, or payment accounts. 

\textbf{G3: Facilitate Admin-Mediated Conflict Resolution Between Buyers and Sellers.}

Another prominent issue was the absence of any structured way to resolve misunderstandings between buyers and sellers. This scenario urges us to design a conflict resolution channel where community admins can act as mediators between the accuser and the accused.

\textbf{G4: Enable Verified Escalation to Formal Authorities.}

Participants explained that some cases of fraud were too severe to be resolved within the community and required intervention from external authorities. However, individual reporting was often ineffective. This scenario urges us to design a system where only community-verified cases with complete evidence can be escalated to relevant authorities, including Police and financial institutions.

\begin{figure}[t]
    \centering
    \includegraphics[width=\linewidth]{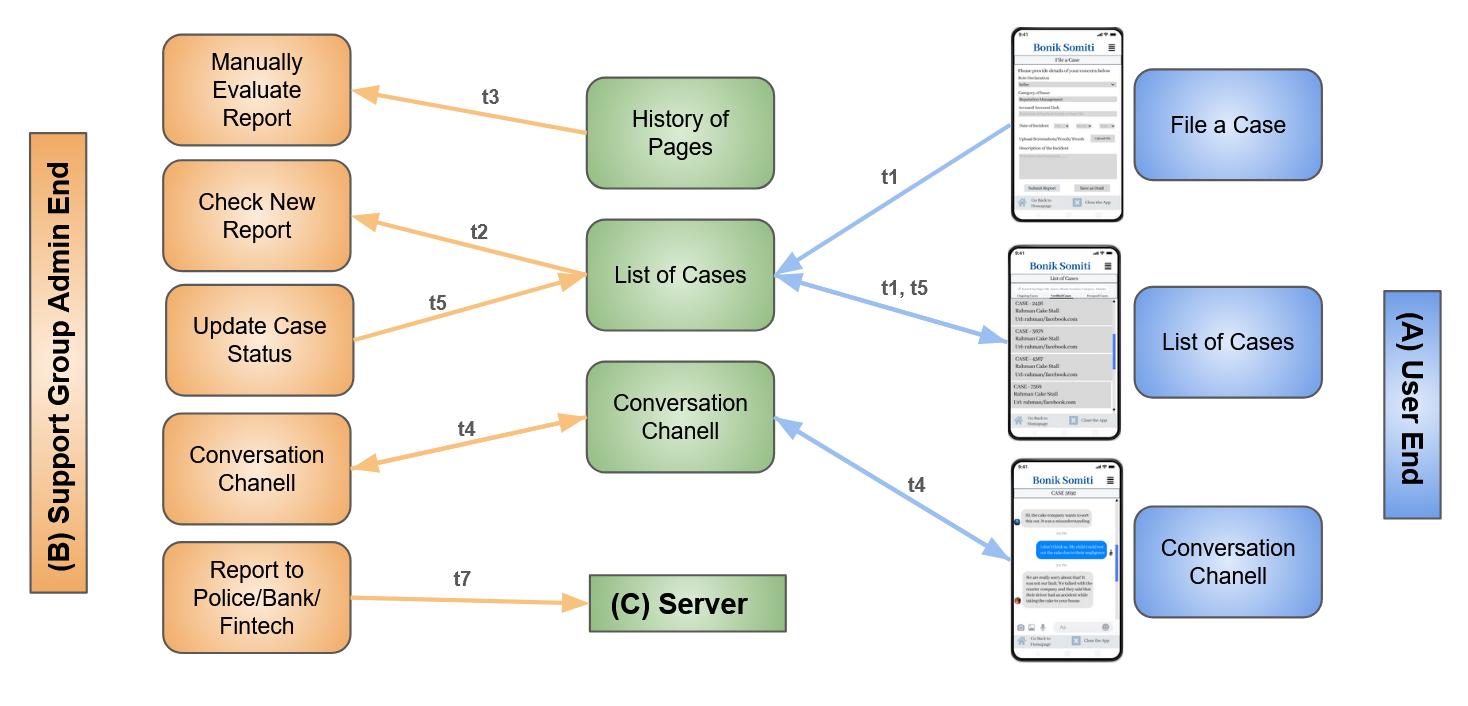}
    \caption{Work-flow diagram of Bonik Somiti. (A) User End lets buyers and sellers file structured cases with role declaration, issue categories, accused account links, descriptions, and evidence, participate in admin-mediated conversation channels, and view lists of ongoing, decided, and dropped cases to support decision-making. (B) Support Group Admin End lets community admins review new reports, manually evaluate submitted evidence using page history records, initiate and moderate conversation channels between accusers and accused parties, update case statuses, and report verified and severe cases to external authorities including law enforcement agencies, banks, and financial service companies; (C) Server holds users’ reports, conversation records, case lists, and historical records of pages and related accounts, and functionalizes the application by connecting the User and Admin ends to support reporting, verification, conflict resolution, and institutional escalation.}
    \label{fig:workflow}
\end{figure}

\subsection{Components and Workflow of Bonik Somiti}

The system has three main components: (A) Users, (B) Support Group Admins, and (C) the Server. Refer to Fig. 1 to follow their details below.

\textit{(A) Users.} Users include both buyers and sellers participating in informal e-markets. To functionalize G1, the system allows users to file a case through a structured reporting interface. While filing a case, users declare their role (buyer or seller), select the category of the issue, provide the accused account link, upload screenshots or other supporting evidence, and describe the incident in detail. Users can also access a List of Cases view, where they can track the status of reported and reviewed cases, including ongoing, decided, and dropped cases. While users cannot modify case outcomes, they can search cases using identifiers such as \textit{Facebook} page names or mobile financial service numbers, helping them make more informed decisions when engaging in future transactions.

\begin{figure}[t]
    \centering
    \includegraphics[width=\linewidth]{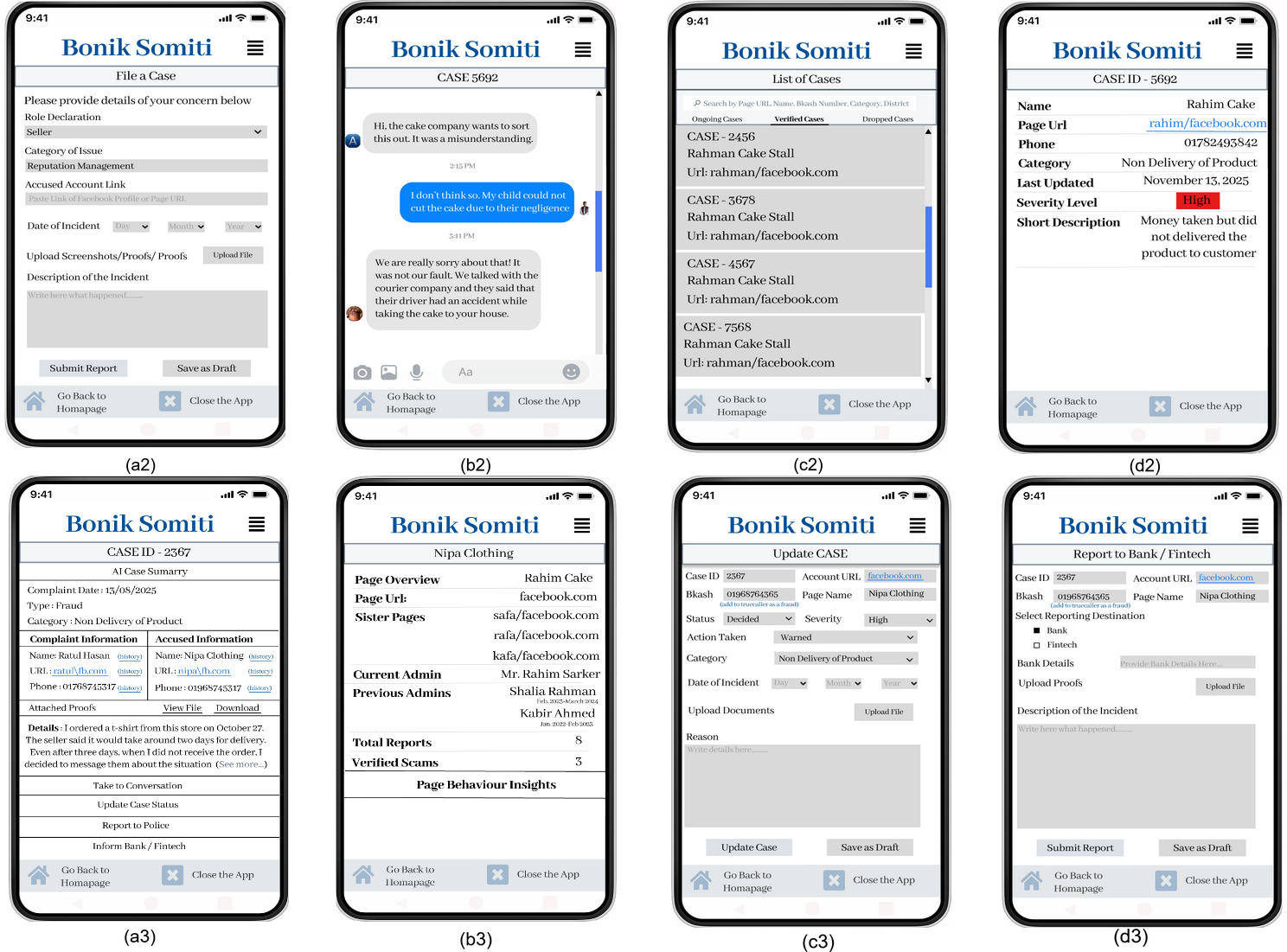}
    \caption{Example pages from Project Bonik Somiti’s Application. (a2–d2) Users’ POV of the accountability interface. (a2) Users file a structured case by submitting role information, issue category, accused account link, descriptions, and supporting documents. (b2) A moderated conversation channel where the accuser, accused, and support group admin communicate to clarify misunderstandings and attempt conflict resolution. (c2) A list of confirmed cases that have been verified by support group admins as potentially risky. (d2) A detailed case view displaying comprehensive information about a selected case, including accuser and accused page names, contact details, evidence, and verification records. (a3–d3) Support Group Admins’ POV of the case management interface. (a3) A consolidated view of newly submitted cases showing complete information about both accusers and accused parties, with options to initiate conversation, report to external authorities, or escalate cases. (b3) Access to the history of pages, allowing admins to review prior reports, linked accounts, phone numbers, and related records to assess patterns of risky behavior. (c3) After manual evaluation and verification, admins update the case status as ongoing, confirmed, or dropped. (d3) For verified and severe cases, admins report evidence-backed case records to banks or financial service providers to enable timely interventions.}
    \label{fig:workflow}
\end{figure}

\textit{(B) Support Group Admins.} In Bonik Somiti, ``support group admins” are verified community moderators, primarily fraud alert group admins, and in some cases experienced buyers and sellers who are selected for moderation responsibility. Once users submit reports, admins can view newly filed cases and manually evaluate the submitted details and evidence. To support G2, admins have exclusive access to the History of Pages, which aggregates records such as previous reports, verification outcomes, related pages, associated phone numbers, linked financial accounts, and AI assisted page behaviour summary. This historical information enables admins to identify repeated or concealed fraudulent behavior and assess the severity of each case.

To functionalize G3, admins can initiate a Conversation Channel between the accuser and the accused when they believe the conflict may be resolved through dialogue. Admins send a conversation request, and the channel is activated only if both parties agree to participate. Through this mediated communication space, admins attempt to clarify misunderstandings and facilitate resolution without escalating the case unnecessarily. 

After reviewing all available evidence and, mediating conversations, admins update the case status. They may drop the case, finalize a verdict, or mark it as ongoing based on their evaluation. To functionalize G4, admins can escalate verified and severe cases to external entities, including law enforcement agencies, banks, or financial service companies. Since these reports are community-verified and evidence-backed, admins can submit complete case records, increasing the likelihood of institutional action.

\textit{(C) Server.} The server stores and manages all system data, including user-submitted reports, case statuses, conversation records, and historical page information. It acts as the central infrastructure connecting users and admins. All reports and updates are routed through the server, which maintains the List of Cases and History of Pages databases. Conversation channels are also hosted on the server to ensure proper documentation and traceability.

%% file: 5.phase-3-method.tex
\section{USERS’ FEEDBACK ON BONIK SOMITI}

In this phase of the study, we sought feedback on our design from the participants. We shared our prototype with users. We first prepared a user-study package, which included a prototype version of Bonik Somiti and a user guide for the application. Some participation happened in person, and some were online, based on the participants’ convenience. During the user study, we showed each step of the prototype to the participants and asked for their feedback.

\subsection{Methods Used to Collect User-Feedback}

\subsubsection{Focus Group Discussions.}

We conducted five FGD sessions with 24 participants. Thirteen participants had previously taken part in our earlier study activities, and eleven were newly recruited. We organized the FGDs by stakeholder group so that participants could speak freely with others who shared similar experiences. We held one FGD with police officers, one with bank and fintech workers, one with fraud-reporting group admins, and two combined FGD with buyers and sellers. We recruited participants by circulating a call for participation through email lists within the authors’ professional and academic networks and by posting the call publicly on social media. The call described the purpose of the work, and interested participants contacted us directly to participate. One FGD (police officers) was conducted in person, while the other four were conducted online over Zoom, based on participants’ preferences. In each session, we explained the purpose of Bonik Somiti, showed participants our prototype, and guided them through the interface step-by-step. We then discussed what benefits such a system might bring for informal e-market communities in Bangladesh. We also discussed potential challenges the system might create for prospective users. All sessions were conducted in Bengali. The average length of each session was 35 minutes. Participation was voluntary and unpaid.

\begin{table}[t]
\centering
\renewcommand{\arraystretch}{1.3}
\setlength{\tabcolsep}{6pt}

\begin{tabular}{|p{1.5cm}|p{3.0cm}|p{4.0cm}|p{4.0cm}|}
\hline

\multicolumn{4}{|c|}{\textbf{Number of interview participants: 8 (Female: 5, Male: 3)}} \\ \hline

\textbf{Age Range (in Years)} &
\textbf{Participant Roles} &
\textbf{Years in Informal E-Market} &
\textbf{Platform Used by Participant} \\ \hline

18--25: 3  \newline
26--35: 4  \newline
36--45: 1  \newline
46+: 0  &
Buyer: 2  \newline
Seller: 3  \newline
Buyer+Seller: 3  \newline
Fraud Group Admin: 0  &
Less than 6 months: 0  \newline
6 months - 1 year: 2  \newline
1 - 3 years: 2  \newline
More than 3 years: 4 &
\textit{Facebook Only}: 7  \newline
\textit{Facebook+Instagram}: 1  \newline
\textit{Facebook+WhatsApp}: 0  \newline
All Three Above: 0 \\ \hline

\multicolumn{4}{|c|}{\textbf{Number of FGD Sessions: 5; Number of Participants: 24 (Female:10, Male:14)}} \\ \hline

\textbf{Age Range (in Years)} &
\textbf{Participant Roles} &
\textbf{Years in Informal E-Market} &
\textbf{Platform Used by Participant} \\ \hline

18--25: 4  \newline
26--35: 12  \newline
36--45: 7 \newline
46+: 1  &
Buyer: 4  \newline
Seller: 1  \newline
Buyer+Seller: 5  \newline
Fraud Group Admin: 5 \newline 
Police Officer: 5  \newline
Bank/Fintech Worker: 4  &
Less than 6 months: 1  \newline
6 months - 1 year: 3 \newline
1 - 3 years: 3  \newline
More than 3 years: 8 &
\textit{Facebook Only}: 13  \newline
\textit{Facebook+Instagram}: 2  \newline
\textit{Facebook+WhatsApp}: 0  \newline
All Three Above: 0 \\ \hline

\end{tabular}

\caption{Demographic Characteristics of the Post-design Interview and Focus Group Discussions Participants}
\label{post-demo}
\end{table}

\subsubsection{Interviews.}

We also conducted eight semi-structured interviews, as these participants requested a one-on-one conversation instead of a group discussion. Four of the interviewees had participated in our earlier study activities, while the remaining four were newly recruited. We recruited interview participants using the same process as the FGDs in this phase. All interviews were conducted online via Zoom, based on participants’ preferences. During each interview, we explained the purpose of Bonik Somiti, showed the prototype (Fig. 3(a–d)), and guided participants through the interface step-by-step. We then discussed what benefits this system might bring for informal e-market communities in Bangladesh and also asked about potential challenges and risks the system might create for prospective users and the broader community. All researchers who worked in the first phase of the study also took part in data collection and analysis in this phase. All interviews were conducted in Bengali and lasted 25–30 minutes.

\subsection{Data Collection and Analysis}

We audio-recorded the FGDs and interviews in this phase (with participants’ permission) and securely stored the recordings for analysis. In total, we collected approximately six hours of audio recordings and produced around 40 pages of transcripts. We first transcribed the recordings in Bengali and then translated the transcripts into English. We analyzed the data using open coding and thematic analysis    \cite{boyatzis1998transforming, strauss1990basics}. Two authors independently read the transcripts and allowed codes to emerge from the data. They then met to compare codes, discuss differences, and merge overlapping interpretations. After the first round, we compiled an initial set of codes and noted recurring patterns across stakeholder groups (e.g., concerns about verification and evidence, risks of retaliation, admin power and bias, workflow feasibility, and institutional constraints). Through multiple rounds of discussion and iterative clustering, we grouped related codes into higher-level themes and organized our findings.

%% file: 6.phase-3-feedback.tex
\section{FINDINGS FROM USERS’ FEEDBACK}

We organize the findings from the the user evaluation in three categories: design concerns and suggestions, accountability and justice Concerns in Bonik Somiti, feedback on user interface and process. We discuss them below.

\subsection{Design Concerns and Suggestions}

We received much feedback regarding our design, and our participants made many suggestions to improve the system so that it can work better in informal e-markets. 

First, in one FGD, a participant pointed out that while submitting report it is easy to post a fake complaint out of personal grudges which can waste time and harm innocent people. In the same discussion, another participant asked whether a filter system could be introduced so that admins do not get overwhelmed by too many reports. They specifically raised the idea of using an AI system to detect and reduce fake reports in advance. 

Second, we discussed our design with five police officers in an FGD. They said even if there are no formal rules in Bangladeshi informal e-markets, they can still provide some informal guidelines and suggested us to create a simple rulebook page following their suggestions. They explained that if users can see the rules, understand the severity of fraud and scams, and learn what is considered serious, they may become cautious in advance. They also extended this guideline idea for admins. One officer suggested building an LLM-based support system for admins where rules, policies, and relevant information would be embedded. Admins can ask question, check rules, violations and what kind of punishment applies to different types of offenses. Another officer agreed and said this could reduce pressure on both admins and the police because it would help admins decide which cases are truly severe and worth forwarding. 

Finally, participant raised concerns about fake admins. In one FGD, a participant explained that a random person could pretend to be an admin and directly message users on \textit{Facebook} or other social media, claiming to be an admin of the app. Such fake admins could misuse this trust and harm users. In the same FGD, another participant suggested that the app we should clearly state that users should not communicate with anyone outside the Bonik Somiti system. They felt this warning could help reduce such risks. 


\subsection{Accountability and Justice Concerns in Bonik Somiti}

During the user evaluation, our participants also discussed and expressed concerns about precarious visibility and retaliation, fragile governance and power in moderation, evidence breakdown, and institutional constraints in informal e-markets. We discuss these themes below.

\subsubsection{Precarious Visibility and Retaliation}

While our prototype aimed to create accountability through reporting and case visibility, our participants warned that visibility can also create new harm. In one FGD, a participant explained that many women run online businesses without the full permission of their husbands or families. For them, being publicly identifiable can be dangerous. As the participant said,

\begin{quote}
\textit{``...If this app exposes their identity, address, or personal information, it could put them at serious risk. If someone posts a fake review or false report against them, it could be very harmful. Because of this fear, many women may also avoid posting complaints or reports themselves, since they do not want their identity to be revealed" \textbf{(P44)}}
\end{quote}


In one FGD participants also debated whether users should be able to see the identities of admins who review cases. One participant suggested that showing admin names would increase trust and accountability. However, others immediately pointed out that naming admins could expose them to personal attacks, pressure, or retaliation. Though Participants wanted transparency but they were also concerned that it could harm admin privacy and safety.

Four participants raised legal and social retaliation concerns. P61 pointed out that an accused person might claim that the reporting process violated their privacy, and then blame our system or the admins. Since this system operates in an informal e-market, he argued that legal boundaries are unclear and could be used as a threat. The accuser could even file a case against the system or the admins if they want.

In addition, 13 participants explained that reporting itself can become a tool of domination. They warned that some buyers may use the system to pressure sellers by threatening to expose their identity, damage their reputation, or file repeated reports. Nine Participants also worried about visiblity of “ongoing cases” section. They explained that sharing an under-review case publicly can damage a seller’s reputation. As P46 emphasized,

\begin{quote}
\textit{``Publicly showing ongoing cases can harm a seller’s business, even if they are innocent. It would be better to hide ongoing cases and only show cases that are verified and proven." \textbf{(P46)}}
\end{quote}

Overall, participants expected privacy-preserving reporting that protects vulnerable sellers, cautious reporters, and even admins. Their feedback suggests that the Bonik Somiti should minimize identity exposure, reduce opportunities for harassment through admin access, and carefully control what is shown publicly during investigations.

\subsubsection{Fragile Governance and Power in Moderation}

While our prototype relied on admins as case evaluators, sixteen participants questioned about the fairness of admins. They explained that the system can easily backfire if admins carry personal grudges, show favoritism, or have conflicts of interest as buyers or sellers. In one FGD, a participant raised concerns about our page history database. She said the database stores personal information, and if admins have bad intentions, they could misuse this information and cause harm. As she said

\begin{quote}
\textit{``The role of admins is very important. If admins hold personal grudges, are biased, or are themselves buyers or sellers with conflicts, the system can backfire. In such cases, your history of page database reports could be misused against certain people. Therefore, who becomes a group admin matters a lot." \textbf{(P42)}}
\end{quote}

In the same FGD, another participant suggested that some data should be deleted regularly to prevent misuse. She also added that there should be a policy to ensure admin accountability.




In another FGD one seller bring out a topic of coordiated fake reporting. Others in the same FGD agreed that and explained that competitors may work together to target a seller by submitting many reports and it will be very difficult to defend against and manage such attacks. Finally, Fourteen participants pointed out that the system may miss a large portion of everyday fraud because many people do not bother to report small losses. As P48 explained

\begin{quote}
\textit{``Not everyone will use the system. Many people are one-time buyers or sellers, and if they lose a small amount of money, they may not consider it serious enough to report. These small scams may seem insignificant individually, but collectively they cause major harm. The challenge is how to bring these small, everyday scams into the reporting system." \textbf{(P48)}}
\end{quote}

Because of this, people who are confident, vocal, or have social power are more likely to report, but many others will stay silent. Due to this the records may not show the real situation of harm in informal e-markets.

\subsubsection{Evidence Breakdown and Institutional Constraints}

While our prototype asked users to submit evidence to support claims, our participants raised concerns about the effectiveness of evidence to resolve disputes in informal e-markets. In one FGD, a group admin created a hypothetical scenario and explained that when using the conversation channel to resolve disputes, some cases can become nearly impossible to judge because both parties can present convincing evidence.

\begin{quote}
\textit{``In some cases, it is impossible to determine who is actually at fault. For example, a buyer receives a parcel, opens it, and finds nothing inside, and records a video. The seller claims the product was in the box and shows their own video proof. In such cases, it is very difficult to decide who is telling the truth." \textbf{(P51)}}
\end{quote}

In the same discussion, another participant agreed with the above point and also warned that evidence can be falsified. He said that in informal e-markets, generating fake proof is very common and it would be hard to distinguish it from the real one.

\begin{quote}
\textit{``Fake photos, fake unboxing videos, fake warranty cards, fake clips are common. Both parties may accuse each other of lying. Determining which evidence is fake will be extremely challenging for admins." \textbf{(P49)}}
\end{quote}

Both of these scenarios show that even if proper evidence exists, it is hard to prove what happened in between and also whether the evidence is fake or not. It would be hard for admins to clearly give a verdict with these challenges.

In the FGD with bank and fintech workers they discussed some limits of financial enforcement. They explained that even if the platform identifies and submits proper evidence, they cannot easily block payments or freeze accounts because they have to follow formal policies. As P58 said

\begin{quote}
\textit{``We cannot freeze an account unless there is instruction from a formal authority. But if your support group admins include police officers or legal authorities, then we can temporarily block transactions. However, for proper implementation, formal legal approval and policy design are required." \textbf{(P58)}}
\end{quote}

In the same FGD, a fintech company admin explained that blocking mobile financial service accounts is often complicated. Because many accounts are shared within families, the registered name and phone number may not always match the actual user. As a result, blocking one account can affect multiple family members and cause unintended harm.

Participants also raised concerns about whether people would use the system at all. In one FGD, P41 explained that many buyers and sellers already rely on \textit{Facebook} for informal buying and selling. They argued that people may not have the motivation to check a separate platform. They stated that if the reporting system could be integrated with \textit{Facebook}, then it would be best.





\subsection{Feedback on User Interface and Process}

During the user evaluation, the participants gave us some feedback on the user interface (UI) and the overall process of the system. First, sixteen participants raised concerns about processing time. They explained that lengthy processing time could be a major barrier for real use. In a similar direction, twelve participants said that ongoing investigations can feel frustrating when users do not know how long a case will take. They suggested showing an estimated time, such as a progress bar or a status indicator in the ongoing case section, so users can understand what is happening and how much time is left. Second, participants discussed deployment and device support. Second, participants suggested small UI changes to improve day-to-day use. In one FGD, P42 suggested adding a “Copy” button in the case summary section so users or admins can easily copy the summary text when needed. In the same discussion, P43 suggested adding an edit option in the Case ID or name field so users can enter and correct their name properly. Third, participants gave feedback on evidence and response workflow. Seven participants said that many transactions and agreements happen through voice messages or phone calls, so the system should allow uploading voice-based evidence as proof. Finally, P47 discussed about the conversation channel and suggested that the accused should have a limited time window to respond and submit counter-evidence. She explained that if the accused does not respond within that time, admins should proceed with a decision so that cases do not remain stuck for too long.


%% file: 7.discussion.tex
\section{Discussion}

This two-phase study investigated fraud, deception, and risk management in Bangladeshi informal e-markets. In Phase-1, our survey and interviews examined the types of scams and harms people encounter, the strategies they use to manage risks and the challanges in combating risks. Based on these findings, we developed design goals to support safer participation in informal e-markets and designed Bonik Somiti. In Phase-2, we conducted user evaluation sessions and gathered feedback from multiple stakeholder groups. Our findings across both phases open discussion on design implications and broader implications for designing safety and accountability tools for informal e-markets.

\subsection{Design Implication}

Our research generates several design implications. A key implication is the need for human–AI collaboration that supports moderation without replacing human judgment. Participants asked for filters to reduce fake reports. Police officers suggested an LLM-based support tool that can help admins quickly check rules, understand possible violations, and decide which cases to forward. At the same time, participants raised concerns about fake evidence. This calls for AI that supports screening, summarizing, and rule lookup, while leaving final decisions to accountable human moderation. This implication extends HCI research on human–AI decision-making by emphasizing support for human decision-making in complex contexts, rather than replacing it \cite{amershi2019guidelines, green2019principles, schulenberg2023towards}. Second, our findings highlight the importance of protecting users from retaliation and misuse of visibility. Participants explained that sharing identity information can be unsafe, especially for women sellers. They also noted that reporting can sometimes be misused to pressure or threaten sellers. This calls for privacy-preserving reporting designs that minimize unnecessary identity exposure and control what is visible during investigations. This implication can connect with broader HCI research on privacy and safety, where visibility, transparency, and accountability must be balanced against harassment, doxxing, and retaliation risks \cite{matthews2025supporting, schoenebeck2023online, marwick2021morally}. Finally, our findings highlight the need for accountable moderation design. Participants argues that the credibility of any reporting system depends on who reviews cases and how decisions are made. Participants raised concerns about biased admins, conflicts of interest, and the misuse of stored information. This calls for governance mechanisms inside the tool, such as clear admin permissions, transparent decision procedures, and routine checks that discourage abuse and favoritism. The system must support not just “moderation,” but also “moderator accountability” as a core design requirement.

\subsection{Broader Implications}

Our research also brings insight into some broader implications. We strongly recommend addressing them while designing tools that aim to support safety, accountability, and risk reduction in informal e-markets.

\subsubsection{Toward Safety and Accountability Beyond Platform Features}

Our research joins the literature on trust and safety in online marketplaces, peer-to-peer commerce, and platform governance, which shows that risk management is not only a technical detection problem but also a social and organizational problem shaped by weak protections and uneven power \cite{luca2017designing, gillespie2018custodians, roberts2019behind}. In Phase-1, participants described frequent fraud and relied on fraud alert groups for safety. However, these groups were fragmented and unreliable. During user feedback, participants warned that Bonik Somiti could face similar risks if if people misuse the reporting feature, if admins receive too many reports, or if people coordinate to submit fake reports. These findings suggest that safety in informal e-markets cannot depend on reporting features alone. It requires governance structures which includes clear procedures for reporting, standards for review, protections against harassment, and accountability mechanisms for those who hold decision-making power \cite{kraut2012building, matias2018civilservant, ma2022m}. This implication invites HCI to treat safety tools as long-term governance systems that must remain fair even under pressure.

\subsubsection{Need for Privacy-Preserving Justice in Informal E-Markets}

Our research contributes to broader discussions in HCI on privacy, visibility, and “justice-making” in socio-technical systems \cite{im2022women, ananny2018seeing}. Many accountability systems assume that visibility and record keeping can reduce harm. Our findings complicate this assumption by showing that visibility can also produce retaliation, gendered harm, and reputational damage. In Phase-1, participants described public shaming and group reporting as common responses. Yet user feedback showed that public visibility can be dangerous, especially for women sellers who may operate without full family support. Participants also warned that showing ongoing cases can destroy a seller’s reputation even if the complaint is false. These concerns align with prior scholarship showing that transparency can increase risk when power is unequal and when people face harassment \cite{sambasivan2018privacy, freed2017digital}. Our findings suggest that justice in informal e-markets must be privacy-preserving and process-based. Instead of designing for maximum exposure, systems should support selective visibility and protections for vulnerable users.

\subsubsection{Toward Institutional Pathways and Policy-Aware Design}

Our work also joins the literature on institutional mediation and socio-legal aspects of digital harm, \cite{nouh2019cybercrime, batool2025between}. Our findings show that institutional pathways are often weak and not clearly applicable in informal e-markets. In our study when fraud occurred participants contacted police, banks, fintech services, but these efforts often did not lead to meaningful resolution. Fintech services cannot freeze accounts without instruction from a formal authority, and proper implementation requires legal approval and policy design. These insights support broader arguments that designing “reporting” systems without institutional alignment can create false expectations of remedy and may even increase harm through misdirected enforcement \cite{nouh2019cybercrime, batool2025between}. This suggests that safety tools for informal e-markets should support realistic escalation workflows and clear communication about what institutions can and cannot do. For HCI, this highlights the need for design work that is aware of policy and institutions. Such work should recognize legal uncertainty, limited enforcement capacity, and the risk of causing harm when punishment is applied in informal settings. \cite{lazar2016human, lynn2025regulating}.

%% file: 8.lim-fw-con.tex
\section{Limitations and Future Work}

Our work has several limitations. First, our participants were recruited through convenience and network-based access, including buyers, sellers, and fraud group admins. Because of this, our findings may reflect the views of people who are already active in online selling and fraud alert communities, and may not represent the full range of informal e-market users across Bangladesh. We therefore treat our findings as exploratory and do not make broad general claims. Second, Bonik Somiti was evaluated as a prototype. In addition, our study could not fully test real integration with platforms like \textit{Facebook} or real enforcement workflows with institutions such as fintech services or police, which participants emphasized as important constraints. In future work, we plan to conduct longer-term field deployment with a wider and more diverse set of participants across different regions, income groups, and digital literacy levels. We will also explore stronger privacy-preserving reporting features, anti-abuse protections  and more robust admin accountability mechanisms. We will pursue collaborations with relevant institutions to design realistic escalation pathways.

